\documentstyle[12pt]{article}
\input epsf
\newcommand{\beq}{\begin{eqnarray}}   
\newcommand{\eeq}{\end{eqnarray}}

\newcommand{\gsim}{\lower.7ex\hbox{$
\;\stackrel{\textstyle>}{\sim}\;$}}
\newcommand{\lsim}{\lower.7ex\hbox{$
\;\stackrel{\textstyle<}{\sim}\;$}}

\input epsf
\newcommand{\grpicture}[1]
{
    \begin{center}
        \epsfxsize=300pt
        \epsfysize=200pt
        \vspace{-5mm}
        \parbox{\epsfxsize}{\epsffile{#1.eps}}
        \vspace{5mm}
    \end{center}
}

\begin{document}
\begin{titlepage}

\begin{flushright}
ITEP-TH-50/97 \\
Saclay/SPhT/T97-116

\end{flushright}
\vspace*{3cm}

\begin{center}
{\Large \bf   Domain Walls  Zoo in Supersymmetric $QCD$}
\vspace{0.8cm}

{\Large  A.V. Smilga
\footnote{ITEP, B. Cheremushkinskaya 25, Moscow 117218, Russia}
$^,$ \footnote{Service de Physique Th\'eorique de Saclay, 91191 Gif--Sur--Yvette,
France} and A.I. Veselov$^{1}$} 

\end{center}

\vspace*{2cm}

\begin{abstract}
Solving numerically the equations of motion for the effective lagrangian
describing supersymmetric $QCD$ with the $SU(2)$ gauge group, we find a
 menagerie of complex domain wall solutions connecting different chirally
 asymmetric vacua. Some of these solutions are BPS-saturated walls; they 
 exist when the mass of the matter fields does not exceed some critical
 value
 $m < m_* = 4.67059\ldots$. There are also sphaleron branches (saddle points
 of the energy functional). In the range $m_* < m < m_{**} \approx 4.83$,
 one of these branches becomes a local minimum (which is {\it not} a BPS
  - saturated one). At $m > m_{**}$, the complex walls disappear altogether
  and only the walls connecting a chirally asymmetric vacuum with the 
chirally
  symmetric one survive.
\end{abstract}

\end{titlepage}

\section{Introduction}

Supersymmetric QCD is the theory involving (in a simplest version) a gauge
 vector supermultiplet
$V$ and a couple of chiral matter supermultiplets $Q^f$ belonging to the
fundamental representation of the gauge group $SU(N)$. The lagrangian
of the model reads
\beq
{\cal L} = \left( \frac{1}{4g^2} \mbox{Tr} \int d^2\theta \ W^2\ + \ {\rm H.c.}
\right)\ + \  \frac{1}{4}\int d^2\theta d^2\bar\theta\ 
\bar S^f  e^V S^f   \nonumber \\
-\left( \frac{m}{4} \int d^2\theta\  S^{f}S_{f} +\mbox{H.c.}\right)\, ,
\label{LSQCD}
\eeq
where $S_f = \epsilon_{fg} S^g$ (for further conveniences, we have 
changed a sign of mass here compared to the standard convention).
The dynamics of this model is in many respects similar to the dynamics
of the standard (non--supersymmetric) QCD and, on the other hand,
supersymmetry allows here to obtain a lot of exact results \cite{brmog}.

Like in the standard QCD, the axial $U_A(1)$ symmetry corresponding to
the chiral rotation of the gluino field and present in the tree--level
lagrangian (\ref{LSQCD}) is broken by anomaly down to $Z_{2N}$. This
discrete chiral symmetry can be further broken spontaneously down to $Z_2$
so that the chiral condensate $<{\rm Tr}\ \{ \lambda^\alpha \lambda_\alpha\}>$ 
($\lambda_\alpha$ is a two - component Weyl spinor describing the gluino
field) is formed. There are
$N$ different vacua with different phases of the condensate
  \beq
  <{\rm Tr}\ \lambda^2> \ =\ \Sigma e^{2\pi i k/N},\ \ \
  \ \ \  k = 0, \ldots, N-1
  \label{cond}
   \eeq
   It was noted recently \cite{Kovner} that on top of $N$ chirally
asymmetric vacua (\ref{cond}), also a chirally symmetric vacuum with
zero value of the condensate exists.

The presence of different degenerate physical vacua in the theory
implies the existence of domain walls --- static field configurations 
depending  only on one spatial coordinate ($z$) which interpolate between
one of the vacua at $ z = -\infty$ and another one at $z = \infty$ and
minimizing the energy functional. As was shown in \cite{Dvali}, in many
cases the energy density of these walls can be found exactly due to the
fact that the walls present the BPS--saturated states. The key
ingredient here is the central extension of the $N=1$ superalgebra
\cite{Dvali,my}
  \beq
\{ \bar Q_{\dot\alpha} \bar Q_{\dot\beta}\}
= 4
\left(\vec\sigma\right)_{\dot\alpha\dot\beta}\int
 \, d^3 x \, \vec\nabla  \left\{\left[
-\frac{m}{2} S^{ f} S_{ f} + \frac{1}{16\pi^2}
\mbox{Tr}\ W^2\right]
- \frac{N}{16\pi^2}
\mbox{Tr}\ W^2 \right\}_{\theta = 0} \, ,
\label{cexm}
  \eeq
A domain wall is a configuration where the integral of the full
derivative in the RHS of Eq.(\ref{cexm}) is non--zero so that the standard
$N=1$ SUSY algebra in the wall sector is modified. A BPS--saturated
wall is a configuration preserving 1/2 of the original supersymmetry i.e.
a configuration annihilated by the action of two certain real linear
combinations of the original complex supercharges $Q_\alpha, 
\bar Q^{\dot{\alpha}}$
\footnote{Such BPS--saturated walls were known earlier in
2--dimensional supersymmetric theories (they are just solitons there)
\cite{WO,Vafa} and were considered also in 4--dimensional theories in
stringy context \cite{Rey}. }.

Combining (\ref{cexm}) with the standard SUSY commutator $\{Q_\alpha,
\bar Q^{\dot\beta}\} = 2 (\sigma_\mu)_\alpha^{\dot\beta} P_\mu$ and bearing
in mind that the vacuum expectation value of the expression in the
square brackets in Eq.(\ref{cexm}) is zero due to Konishi anomaly
\cite{Konishi} 
\footnote{See recent \cite{Chib} for detailed pedagogical explanations.},
it is not difficult to show that the energy density of a BPS--saturated
wall in SQCD satisfies a relation
  \beq
   \label{eps}
   \epsilon \ =\ \frac N{8\pi^2} \left|<{\rm Tr}\ \lambda^2>_\infty
    \ -\ <{\rm Tr}\ \lambda^2>_{-\infty} \right|
    \eeq
where the subscript $\pm \infty$ marks the values of the gluino
condensate at spatial infinities. The RHS of Eq.(\ref{eps}) presents an 
absolute   lower bound for the energy of any field configuration interpolating
between different vacua.
    
    The relation (\ref{eps}) is valid {\it assuming} that the wall is
BPS--saturated. However, whether such a BPS--saturated domain wall
exists or not is a non--trivial dynamic question which can be answered
only in a specific study of a particular theory in interest. 

In Refs.\cite{my,SV} this question was studied for the $SU(2)$ gauge group
in the framework of the effective low energy lagrangian due to Taylor,
Veneziano, and Yankielowicz \cite{TVY}. 
It is written in terms of the composite colorless chiral superfields
  \beq
  \label{norm}
\Phi^3 = \frac 3{32\pi^2} {\rm Tr}\ \{W^2\}, \ \ X^2 = S^fS_f
  \eeq
 The effective
{\it potential} is rigidly fixed from
the requirement that the conformal and the chiral anomaly of the 
theory (\ref{LSQCD}) under consideration are reproduced correctly.
There is some freedom of choice for the kinetic term which we will discuss
 later in the paper.
The effective TVY lagrangian presents a Wess--Zumino model 
 with the superpotential    
\beq
{\cal W} = \frac{2}{3} \Phi^3 \left[ \ln \frac{\Phi^3 X^2}
{\Lambda_{SQCD}^5} \ -\ 1 \right] - \frac{m}{2} X^2 
\label{WTVY}
\eeq
The corresponding potential for the lowest components $\phi$, $\chi$ of
the superfields $\Phi$, $X$
 \beq
U(\varphi, \chi) \ =\ \left|\frac{\partial {\cal W}}{\partial \varphi}\right|^2 
+
 \left|\frac{\partial {\cal W}}{\partial \chi}\right|^2 \ =\ 
4\left| \varphi^2 \ln(\varphi^3 \chi^2) \right|^2 +
\left|\chi\left(m - \frac{4\varphi^3}{3\chi^2} \right)\right|^2
\label{potTVY}
  \eeq
( from now on, with few exceptions, we will measure
everything in the units of $\Lambda_{SQCD} \equiv \Lambda$) has three 
non--trivial
minima:
  \beq
\phi\ =\ \chi \ =\ 0;
\label{min0}
 \eeq
   \beq
\phi \ =\ \left( \frac {3m}4 \right)^{1/6}, \ \ \ \chi \ =\ 
\left(\frac 4{3m} \right)^{1/4}\ ; \nonumber \\     
\phi \ =\ e^{-i\pi/3}\left( \frac {3m}4 \right)^{1/6}, \ \ \ \chi \ =\ 
i\left(\frac 4{3m} \right)^{1/4} 
 \label{minn0}
  \eeq
There are also the minima with the inverse sign of  $\chi$ (and the
appropriately chosen phase of $\phi$), but
they are physically the same as the minima (\ref{minn0}): the vacuum
expectation values of the gauge invariant operators $<{\rm Tr}\ \lambda^2>
\ = \ (32\pi^2/3) <\phi^3>$ and $< s^f s_f> \ =\ <\chi^2>$
($s^f$ is the squark field) are the same. The vacua (\ref{minn0})
 involve a non-zero gluino condensate (\ref{cond}). The vacuum (\ref{min0})
 is chirally symmetric.

To study the domain wall configurations, we should add to the potential
(\ref{potTVY}) the kinetic term which we choose in the simplest possible
form ${\cal L}_{\rm kin} \ =\ |\partial \phi|^2 
+ |\partial \chi|^2$  and solve the equations of motion
with appropriate boundary conditions. In Ref.\cite{my}, it was done for the
walls connecting a chirally asymmetric vacuum with the chirally symmetric
one. It turned out that the solution exists for any value of $m$ and that
it is always BPS -- saturated, i.e. the walls could be found solving
not the equations of motion, but the more simple first order BPS equations
(we again refer  a reader to Ref.\cite{Chib} for details)
  \beq
  \partial_z \phi \ =\ \pm \partial \bar {\cal W} /\partial \bar \phi,
  \ \ \ \ \      \partial_z \chi \ =\ \pm \partial \bar {\cal W} /\partial
\bar \chi\ 
  \label{BPS}
  \eeq
The surface energy density of these walls coincides with its lower bound 
(\ref{eps}). Explicitly, in physical units
 \beq
 \label{ereal}
 \epsilon_0 \ =\ 4 \sqrt{(m\Lambda_{SQCD}^5)/3}
  \eeq

However, there is also a different kind of walls which connect different
chirally asymmetric vacua in (\ref{minn0})
\footnote{We will call these walls {\it complex walls}
in contrast to the {\it real walls} studied in Ref.\cite{my}.
For the walls separating the symmetric and an asymmetric vacua, the phase
of the fields does not change.
It can be set to zero which considerably simplifies the equations.
 For the walls separating different vacua
in (\ref{minn0}), the trajectories $\phi(z)$ and $\chi(z)$ are essentially
complex, and the problem is technically more intricate.}.
The BPS equations with the corresponding boundary conditions were solved
numerically in \cite{SV}. It was shown that the solutions exist only
in the limited range of masses $m \leq m_* = 4.6705\ldots$. At larger
masses, complex BPS domain walls {\it do} not exist. A kind of phase
 transition occurs.

In this paper, we perform a more detailed study of the complex walls. We show
that on top of the {\it upper} BPS branch found in Ref.\cite{my}, also a 
{\it lower} BPS branch exists. Here the terms ``upper'' and ``lower'' refer
not to energy (  the energy of both branches is, of course, 
the same and is given by the
BPS bound $2\epsilon_0$), but to some other parameter: the value of $|\phi|$
at the middle of the wall. When the mass is small, the trajectories can be
found analytically. For the lower branch, the fields in the middle of the
wall are very small in this case so that the wall presents a very loose bound
state of two real walls. When the mass rises, $|\phi|_{\rm middle}$ also 
rises. At $m = m_*$, two branches fuse together.

We also studied numerically the wall solutions of the second order equations
of motion. Some of the solutions are the BPS walls described above, but some 
are not. At $m < m_*$, there are two new {\it sphaleron} branches. They 
present saddle points of the energy functional. The upper (in the
 same sense as above)  sphaleron branch presents a ``mountain pass'' 
 between two different
 BPS valleys. The lower sphaleron  branch presents a mountain pass
 between the lower BPS valley and the field configuration corresponding
 to two real BPS walls separated at infinite distance. At $m=m_*$ when 
 two BPS valleys fuse together, the upper sphaleron solution coincides
 with the BPS one: a mountain pass disappears and there is only one local
 minimum of the energy functional. This local minimum (and the corresponding
 solution for the equations of motion) exists also at somewhat larger masses,
 in the range $m_* < m < m_{**}$. It is not, however, a BPS - saturated 
minimum
 anymore, its energy somewhat exceeding the BPS bound. At 
 $m = m_* \approx 4.83$,
 the upper branch fuses together with the lower sphaleron branch. At larger
 masses, {\it no complex domain wall solutions exist}.
\footnote{To avoid misunderstanding, we emphasize that two chirally asymmetric
vacua can always be connected in two steps, via the symmetric one. That is
 there is always a ``composite''
 solution 
presenting a combination of two real walls separated at infinite distance.}

 The plan of the paper is the following. Before proceeding with
 numerics, we discuss in the next section the limits $m \to 0$ and
 $m \to \infty$ where the problem can be studied analytically. In Sect.3, 
 we describe in
 details the numerical solutions of the BPS equations in the intermediate
 mass range.  In Sect.4, we solve 
 the second order equations of motion and display the existence of two 
 sphaleron branches.  The 
 last section is devoted to the discussion of  physical relevance of our
 results and to some speculative remarks on the issues which are not yet
clear.

\section{Exploring the limits.}
\setcounter{equation}0

The situation is greatly
simplified when the mass parameter $m$ in the lagrangian (\ref{LSQCD}) is
large or small (compared to $ \Lambda $). 
Consider first the case of large masses. In this case, one can integrate
the heavy matter fields out and write the effective lagrangian for the
composite chiral superfield $\Phi$. Technically, one should use the
 Born--Oppenheimer procedure and to freeze down the matter field $\chi$ so
 that the large second term in the potential (\ref{potTVY}) disappear.
 Proceeding in supersymmetric way, we get $X^2 = 4\Phi^3/3m$. Substituting
 it in the first term, we obtain the Veneziano--Yankielowicz effective
 lagrangian \cite{VY} which is the Wess--Zumino model for the single chiral 
 superfield $\Phi$ with the superpotential 
  \beq
  \label{WVY}
  {\cal W} \ =\ \frac 43 \Phi^3 \left[ \ln \frac {\Phi^3}{\Lambda_{SYM}^3}
  -1 \right]
  \eeq 
where $\Lambda_{SYM}^3 = \sqrt{3m \Lambda_{SQCD}^5/4}$. The expression
(\ref{WVY}) and the corresponding expression for the potential
  \beq
  \label{potVY}
  U(\phi) =  |\partial {\cal W}/\partial \phi|^2 = {16} |\phi^4|
|\ln \phi^3|^2
  \eeq
  (here $\phi$ is measured in the units of $\Lambda_{SYM}$)
 is not yet well defined: the logarithm has many sheets, and one should specify
 first what particular sheet should be taken. An accurate
 analysis taking into account the fact that the topological charge 
   \beq
  \label{nu}
\nu \ =\ \frac 1{16\pi^2}
 \int {\rm Tr} \{G^2 \tilde G^2 \} d^4x
  \eeq
 in the original theory
(\ref{LSQCD}) is quantized to be integer reveals that the true potential
is glued out of two pieces \cite{Kovner,my}:
   \beq
   \label{glue}
 \ln \phi^3 \equiv \ln|\phi^3| + i {\rm arg} (\phi^3), \ \ \ \ 
 {\rm arg} (\phi^3)\  \in \ (-\pi/2, \pi/2)
 \nonumber \\
 \ln \phi^3 \equiv \ln|\phi^3| + i [{\rm arg} (\phi^3) - \pi], \ \ \ \ 
 {\rm arg}( \phi^3)\  \in  \  (\pi/2, 3\pi/2)
  \eeq
  (remind that only the field $\phi^3 \sim {\rm Tr}\ \lambda^2$, not
  $\phi$ itself has a direct physical meaning). Quite an analogous situation
  holds in the Schwinger model: the true bosonized lagrangian (in
   the Schwinger model it is just {\it equivalent} to the original theory )
   is glued out of several branches when taking into account the effects
   due to quantization of topological charge \cite{QCD2}.
The potential (\ref{potVY}) has three degenerate minima. Two of them 
with $<{\rm Tr}\ \lambda^2>\  \propto\  <\phi^3>\  \ = \ \pm 1$ are chirally
 asymmetric and the third one with $<\phi>\ =\ 0$ is chirally symmetric.

The domain walls in this model were studied in \cite{my}. It turned out that
only the real domain walls interpolating between a chirally asymmetric and 
the
chirally symmetric vacua exist in this case. Nontrivial complex domain wall
solutions which do not go through zero are absent.
Speaking of the real walls, they are BPS saturated and can be obtained
 solving
the first order equations $\partial \phi = \pm \partial {\cal W}/\partial 
\phi$. The solution is ``almost'' analytical, it is expressed via integral 
logarithms.

Consider now the case of small masses.
In this case, chirally
asymmetric vacua are characterized by large values of the matter scalar
field $\chi$ . 
Again,  the theory involves two different energy scales, and one
can tentatively integrate out heavy fields  and to write the Wilsonean
effective lagrangian describing only light degrees of freedom. Proceeding
in the Born--Oppenheimer spirit, we should freeze now the heavy field 
$\phi$ in the potential (\ref{potTVY}) so that the large first term in the
potential acquire its minimum (the zero) value. In contrast to the
large mass situation, this can now be achieved in two ways: either
by setting $\phi = 0$ or by setting $\phi^3 \chi^2 = 1$. In the first case,
we will obtain the effective lagrangian describing the dynamics of the 
chirally symmetric phase (it is just the lagrangian of free light chiral
field $X$ involving $\chi$  and its superpartner).

 The second choice results
in the lagrangian describing the dynamics of the chirally asymmetric phases.
It is
the lagrangian of the Wess--Zumino model with a single chiral
superfield $X$ and a non-trivial superpotential
  \beq
{\cal W} = -\frac{2}{3X^2} -\frac{m}{2}X^2
\, .
\label{WHiggs}
\eeq
This lagrangian is well known and was obtained earlier from  
instanton and/or from holomorphy
considerations \cite{brmog}.
The corresponding potential $U \ =\ |\partial W/\partial \chi |^2$ has
two different non--trivial minima at
   \beq
\label{chi*}
<\chi^2> \ =\ \pm \chi_*^2 \ = \ 
\pm \sqrt{4/3m}
  \eeq
When $m \ll 1$, large expectation value (\ref{chi*}) results in breaking down 
the gauge symmetry of the original theory by the Higgs mechanism: 
the theory is in the Higgs phase.

 A domain wall interpolating
between two  vacua (\ref{chi*})  is BPS--saturated. The solution can be found
analytically \cite{my}
  \beq
\label{wallhgs}
\chi(z) \ =\ \chi_* \frac{1 + i e^{4m(z-z_0)}}{\sqrt{1 + e^{8m(z-z_0)}}}
  \eeq
where $z_0$ is the position of the wall center. We see that $|\chi(z)|$
stays constant and only the phase is changed.

In this approach, we are not able, however,  to study the real domain walls
interpolating between the
chirally symmetric and a chirally asymmetric vacua. Such a wall corresponds
to going through a high energy barrier separating two kind of vacua. It is
a remarkable consequence of supersymmetry and of the related BPS condition
(\ref{eps}) that the energy of such a wall is still not large. 

It was shown
in \cite{my} that in the limit $m \to 0$, the real wall trajectory can also
be found analytically. In a funny way, the transition from the chirally 
symmetric
to a chirally asymmetric vacuum goes in two stages. On the
first stage, only the field $|\chi|$ is changed rising from  zero at 
$z = -\infty$  to  its maximal value $\chi_*$ at $z =0$,  
 $\chi(z) = \chi_* e^{mz}$, while 
the field $\phi$ stays frozen at zero.
This first
part of the wall has the width $\sim 1/m$ and carries the half of the wall
total energy. At $z > 0$, 
the trajectory abruptly turns. $\chi(z)$ is not changed anymore, but 
$\phi(z)$ changes until it levels off at its  value in the
asymmetric vacuum state $\phi_* = \chi_*^{-2/3}$ . 
The equations for this second stretch have exactly the same form as the BPS 
equations in the pure supersymmetric Yang--Mills theory. The second part
of the wall carries another half of the total energy, but 
is much thinner than the first part, its width being of order
$m^{-1/6}$. It is on this second stage when the potential barrier between
symmetric and asymmetric vacua is penetrated.

Thereby, the physical situation in the limits $m \to 0$ and $m \to \infty$
is somewhat different. In both cases, we have 3 different vacua. However,
an analog of nontrivial complex walls (\ref{wallhgs}) connecting different
asymmetric vacua which are present at small masses, are absent when the mass
is large. In the latter case, only real walls are present. It is therefore
very interesting to understand what happens in between, at intermediate
 values
of masses, and how the transition from one regime to another  occurs.
That was the main motivation for our study.

 Generally, one should study the theory with the potential
(\ref{potTVY}). The status of this effective theory is somewhat more
uncertain than that of (\ref{WHiggs}) --- for general value of mass,
the TVY effective lagrangian is not Wilsonean; light and heavy degrees
of freedom are not nicely separated. But it possesses all the relevant
symmetries of the original theory 
 and satisfies the anomalous Ward identities for correlators at zero
momenta. We think that the use of the TVY lagrangian is justified as
far as the vacuum structure of the theory is concerned.

 \section{Solving BPS equations.}
 \setcounter{equation}0
 Let us choose the positive sign in Eq.(\ref{BPS}) and try to solve it with 
 the boundary conditions
 \beq
 \label{bcwall}
 \phi(-\infty) = \left(\frac{3m}4\right)^{1/6}, \ 
 \phi(\infty)  = e^{-i\pi/3} 
 \left(\frac{3m}4\right)^{1/6}, \nonumber \\  
 \chi(-\infty)  = \left(\frac 4{3m}\right)^{1/4}, \  \chi(\infty)  = i 
 \left(\frac 4{3m}\right)^{1/4} 
 \eeq
 (the negative sign in (\ref{BPS}) would describe the wall going in
  the opposite direction in $z$). The solution of the equations (\ref{BPS})
   with the boundary conditions (\ref{bcwall}) has the fixed energy density
   which due to  (\ref{eps}) is twice as large as the energy density of the
   real wall (\ref{ereal}). Thereby a complex BPS wall may be thought of as
   a marginary stable bound state of two real walls.

 Technically, it is convenient to introduce the polar variables 
 $\chi = \rho e^{i\alpha}, \ \ \phi = R e^{i\beta}$.  Then the system 
 (\ref{BPS}) (with the positive sign chosen) can be written in the form
   \beq
   \label{4sys}
   \left\{ \begin{array}{l}
\partial_z \rho \ =\ -m\rho \cos (2\alpha) + \frac{4R^3}{3\rho} \cos (3\beta)
\\
\partial_z \alpha \ =\ m \sin (2\alpha) -  \frac{4R^3}{3\rho^2} \sin (3\beta)
 \\
\partial_z R \ =\ 2R^2 \left[ \cos(3\beta) \ln(R^3\rho^2) -
\sin (3\beta) (3\beta + 2\alpha) \right]  \\
\partial_z \beta \ =\ -2R \left[ \sin(3\beta) \ln(R^3\rho^2) +
\cos (3\beta) (3\beta + 2\alpha) \right] 
\end{array} \right.
   \eeq
The wall solution should be symmetric with respect to its center. 
Let us seek
for the solution centered at $z=0$ so that 
\beq
\label{sym}
\rho(z)  = \rho(-z), \ R(z) = R(-z),\ \alpha(z)  = \pi/2 - \alpha(-z),
\ \beta(z)  = -\pi/3 - \beta(-z)
\eeq
Indeed, one can be easily convinced that the Ansatz (\ref{sym}) goes through
the equations (\ref{4sys}). 
\footnote{A logical possibility could be that
 symmetric equations have
a couple of asymmetric solutions. We searched for such but did not
find any, however.}

The system (\ref{BPS}) has one integral of motion \cite{Chib}:
\beq
\label{ImW} 
{\rm Im}\ {\cal W}(\phi, \chi) \ =\ {\rm const}
\eeq
Indeed, we have
$$
\partial_z {\cal W} = \frac{\partial {\cal W}}{\partial \phi} \partial_z \phi  \ + \ 
\frac{\partial {\cal W}}{\partial \chi} \partial_z \chi \ =  \\ 
\left| \frac{\partial {\cal W}}{\partial \phi} \right |^2 \ +\ 
\left| \frac{\partial {\cal W}}{\partial \chi} \right|^2 \ =\ 
\partial_z \bar {\cal W}
$$
It is convenient to solve the equations (\ref{4sys}) numerically on the
half--interval from $z=0$ to $z = \infty$. The symmetry (\ref{sym}) 
dictates $\alpha(0) = \pi/4,\ \beta(0) = -\pi/6$. Then the condition 
(\ref{ImW})
[in our case ${\rm Im}\ {\cal W}(\phi, \chi) = 0$ due to the boundary conditions
(\ref{bcwall})] implies
  \beq
  \label{Rrho}
\frac{4R(0)^3}3 \left\{ \ln[R(0)^3\rho(0)^2] - 1\right \} + m\rho(0)^2 \ =\ 0  
   \eeq
Thus, only one parameter at $z=0$ [say, $R(0)$] is left free. We should fit
it so that the solution would approach the complex minimum in
 Eq.(\ref{minn0})
at $z \to \infty$.

\begin{figure}
\grpicture{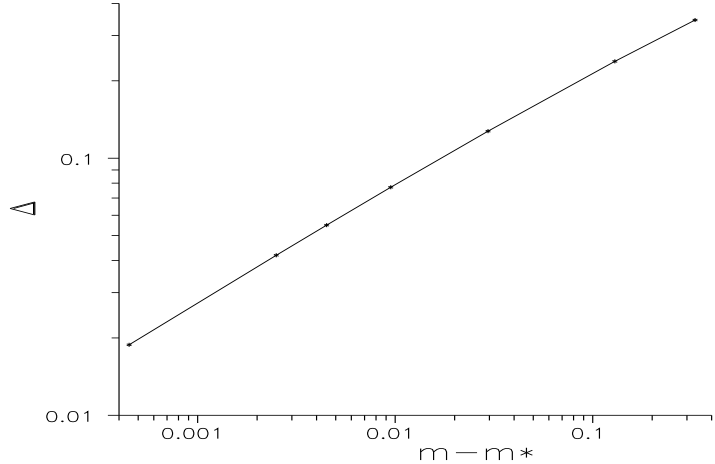}
\caption{ Mismatch parameter $\Delta$ as a function of mass}
\label{mmtch}
\end{figure}

It turns out that the solution of this problem exists, but only in the 
limited
range of $m$. If $m > m_* = 4.67059\ldots$, the solution {\it misses} the 
minimum no matter what the value of $R(0)$ is chosen. This is illustrated in 
Fig. \ref{mmtch} where the ``mismatch parameter''
\beq
\Delta \ =\  \min_{R(0)}  \min_z 
\sqrt{ \left|\chi(z) -  i(4/3m)^{1/4}\right|^2 + 
\left|\phi(z) - e^{-i\pi/3}(3m/4)^{1/6}\right|^2}
\label{mism}
\eeq
 is plotted (in the double logarithmic scale) as a function of mass.
 The dependence $\Delta(m)$ fits nicely the law
  \beq
  \label{DelBPS}
  \Delta_{\rm BPS}(m) \ =\ 0.56 (m - m_*)^{0.44}
  \eeq
It smells like a critical behavior but, as $\Delta$ is not a physical
 quantity,
we would not elaborate this point further.

\begin{figure}
\grpicture{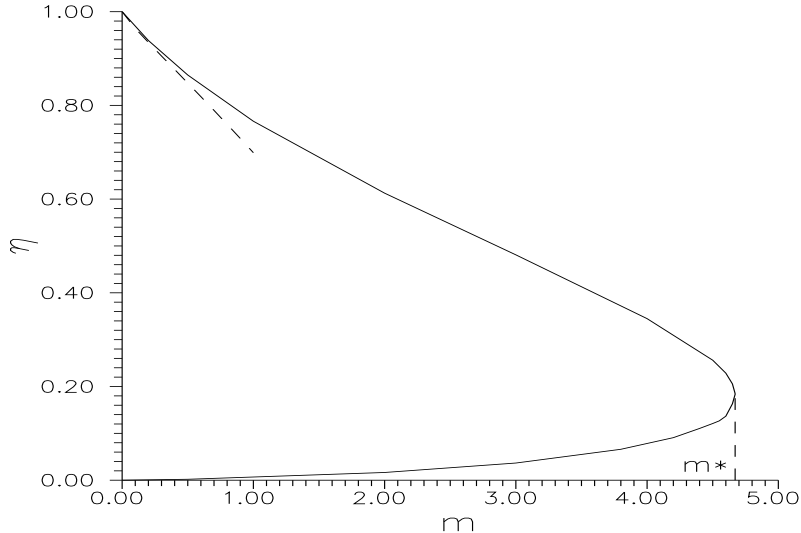}
\caption{The ratio $\eta = R(0)/R(\infty)$  as a function of mass. The dashed
line describes the analytic result (\ref{etam}) valid for small masses.}
\label{etas}
\end{figure}

 At $m = 4.67059$ or at smaller values of mass, the solution exists, however.
Moreover, it turns out that, at $m < m_*$, there are {\it two}
 different solutions:
the upper branch with the larger value of $R(0)$ and the lower branch with the
lower value. 
In Fig. \ref{etas} we plotted the dependence of $\eta = R(0)/R(\infty)$
 on $m$  for both branches. 
 We see that, at $m = m_*$, two branches are joined together. This {\it is}
 the reason why no solution exists at larger masses.

\begin{figure}
\grpicture{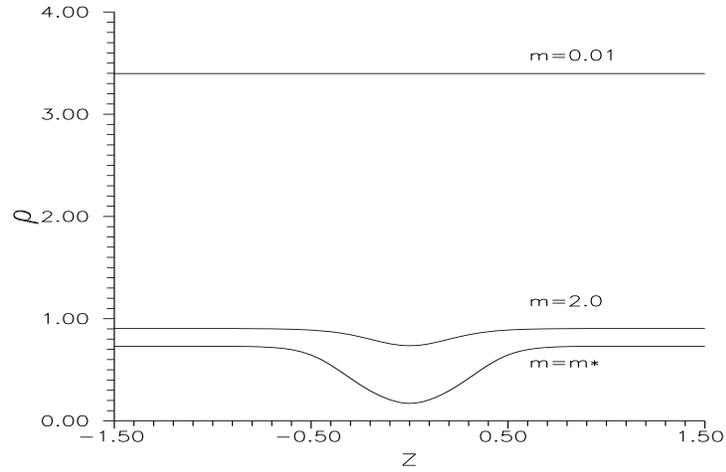}
\caption{ Upper BPS branch: $\rho(z)$ for different  masses.}
\label{rhos}
\end{figure}

\begin{figure}
\grpicture{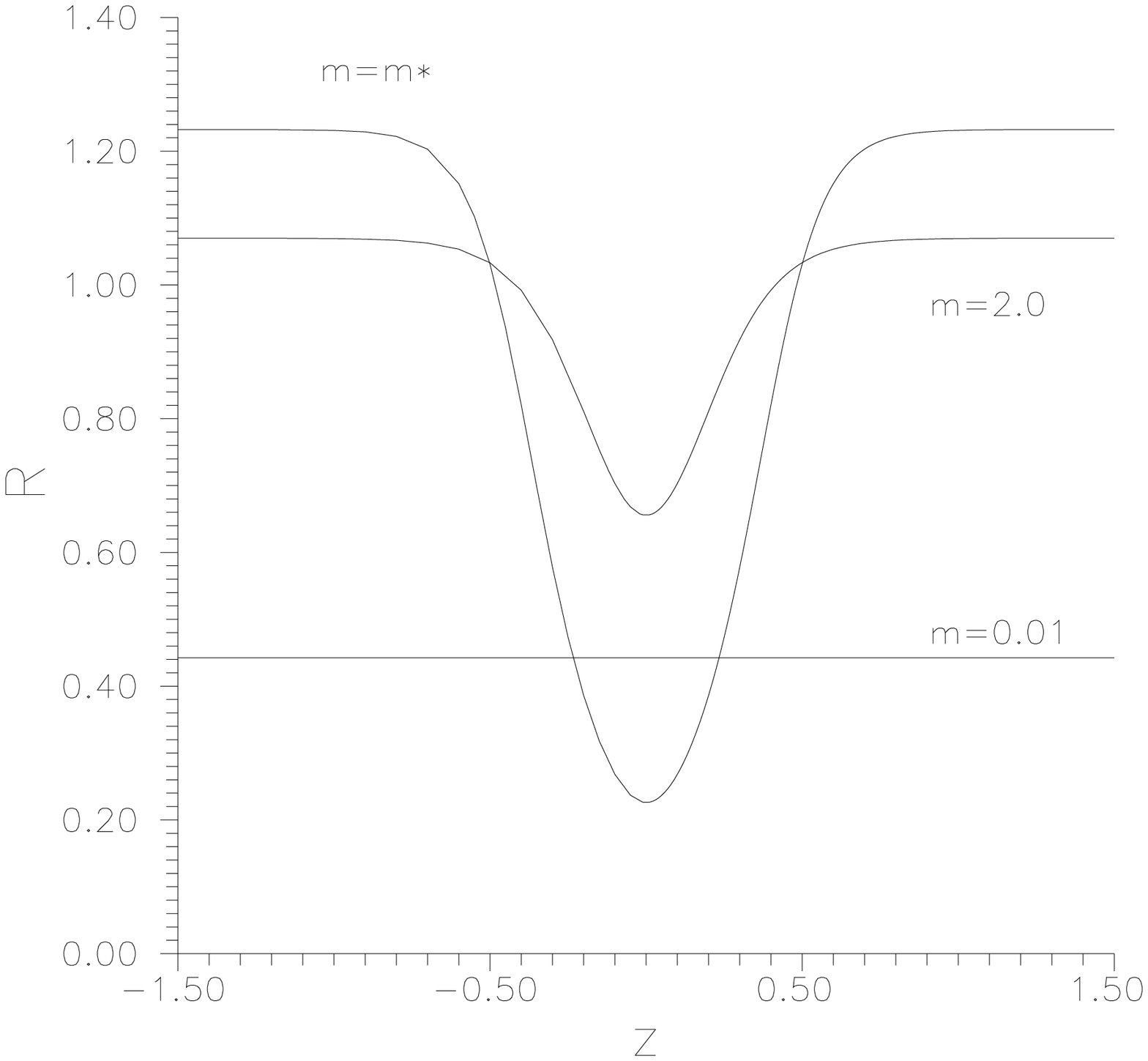}
\caption{Upper BPS branch: $R(z)$ for different masses.}
\label{Rs}
\end{figure}

 Let us discuss
 now what happens with these two branches in the small mass limit. Consider
 first the upper branch. 
 The profiles of the functions $\rho(z)$ and $R(z)$ for three
 values of mass: $m = 0.01,\ m = 2.0,$ and $m = 4.67059$ for the upper
 branch in the whole  interval $-\infty <
 z < \infty$  are plotted in Figs. \ref{rhos},\ref{Rs}. We see that, for small
 masses, 
 $R(z)$ and $\rho(z)$ are virtually constant. This is not surprising, of
 course. For small masses,
  the solution approaches, as it should, the analytic solution 
 (\ref{wallhgs}) (with $\phi \ =\ \chi^{-2/3}$).  

Linearizing at small nonzero masses the equations (\ref{4sys}) with respect 
to small deviations of the
solutions $\Delta \rho(z), \Delta \alpha(z)$, etc. from the analytic
result (\ref{wallhgs}) ( and $\phi = \chi^{-2/3}$)  valid in the
limit $m \to 0$, 
it is possible to determine the dependence $\eta(m)$ in the 
small mass region. The derivation in the general $SU(N)$ case will be 
presented in \cite{SU3}.  Here we
only quote the result for $SU(2)$ :
  \beq
  \label{etam}
  \eta(m) \ =\ 1 - \frac 29 \left(\frac {4m^5}3 \right)^{1/6} - \frac 5{81}
  \left(\frac {4m^5}3 \right)^{1/3} + O(m^{5/2})
  \eeq
  The numerical results for $\eta(m)$ are in the excellent agreement with 
  this formula.

 \begin{figure}
\grpicture{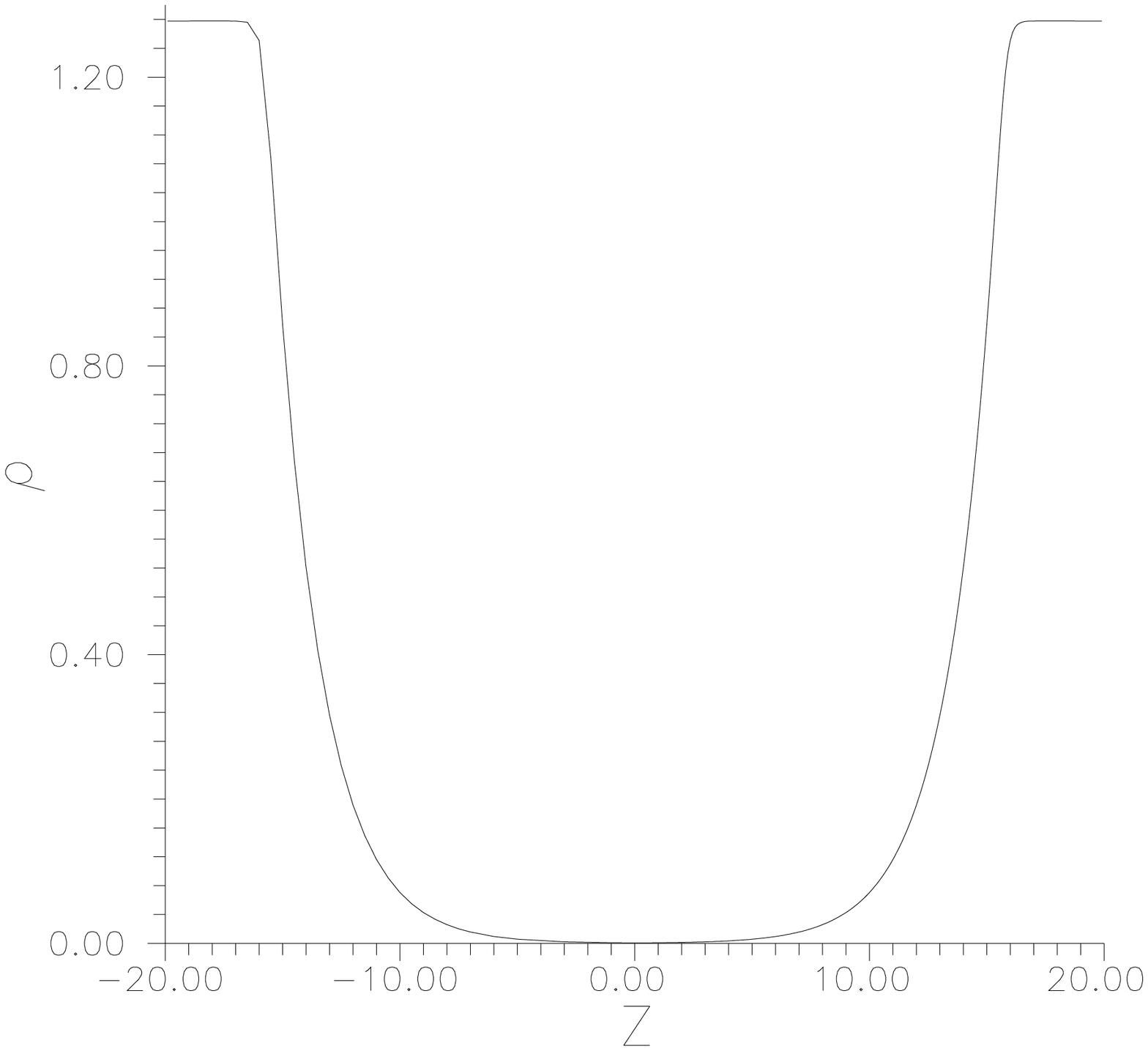}
\caption{Profile $\rho(z)$ of the lower BPS branch at $m=0.5$}
\label{lrhos}
\end{figure}

Consider now the lower branch. $R_{\rm low}(0)$ and, due to Eq.(\ref{Rrho}),
 also
 $\rho_{\rm low}(0)$ go to zero in the limit $m \to 0$. Thereby, the
solution looks very much similar to a combination of two real walls separated
at a large distance (The phases $\alpha(z)$ and $\beta(z)$ are changed, of
course, so that the whole solution is complex, but the change occurs in the 
central region where the absolute values of the fields $R$ and $\rho$ are
very small.).
One can say that it is a very loosely bound state of two real
walls. The profile $\rho(z)$ for the lower BPS branch at $m = .5$
found numerically is shown in Fig. \ref{lrhos}. 
Indeed, the complex wall is  split apart in two widely separated real 
walls. $\rho(0) \approx \ 6.7 \cdot 10^{-4}$ in this case. We observe 
numerically that, for small masses,  $\rho(0)$ and $R(0)$ go down $\propto
\exp\{-C/m^\alpha\}$ with $\alpha \approx 1/4$, $C_\rho \approx 6.1$, 
$C_R \approx 5.3$. 

The revealed phenomenon --- the appearance of two solutions of the equation
system (\ref{4sys}) at $m < m_*$ and the absence of solutions at $m > m_*$
can be understood in the framework of the qualitative theory of differential
equations (see Ref.\cite{Vafa} for a related discussion). The phase space
of the system (\ref{4sys}) is 4--dimensional. The space of all possible 
integral
curves is 3--dimensional (there are 4 integration constants, but one of them
corresponds to  shifting  the independent variable $z$ along the trajectory).
 
The system involves
the integral of motion (\ref{ImW}) so that the whole 3--dimensional space
of the trajectories presents a set of 2--dimensional slices with a given
value of Im ${\cal W}$.

We are interested in a special trajectory belonging to the slice with Im 
${\cal W} \ = \ 0$ and passing through two chirally asymmetric minima
 (\ref{minn0}).
These minima present singular points of the equation system (\ref{4sys})
where the right hand sides of the equations turn to zero. Thereby, the alias
for our domain walls is {\it separatrices}. 

Linearizing the equations at the vicinity of singular points and studying
the matrix of the linear system thus obtained (the Jacobi matrix), one can
be convinced that these points have saddle nature: the Jacobi matrix has 
two positive $\lambda$, $\mu$ and two negative $-\lambda$, $-\mu$
 eigenvalues.
The space of all possible trajectories which have their origin at one of the
singular points (say, the point with real $\phi$, $\chi$) is 1--dimensional
--- only the directions in the phase space along the eigenvectors of the
Jacobi matrix with {\it positive} eigenvalues should be taken into account.
There are two such eigenvectors which can be mixed with 1 real parameter.
Likewise,  
 the subspace of the
trajectories which end up at the point $\phi = e^{-i\pi/3} (3m/4)^{1/6},
\ \chi = i(4/3m)^{1/4}$ is also 1--dimensional. 
They correspond to a mixture of two eigenvectors of the corresponding Jacobi
matrix with {\it negative} eigenvalues.
The wall solution belongs
 to both subspaces.

\begin{figure}
\grpicture{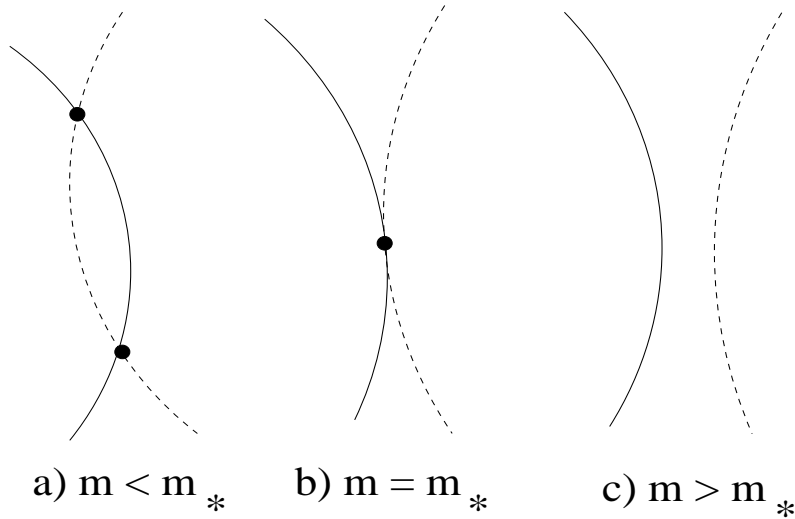}
\caption{A space of trajectories with ${\rm Im}\ {\cal W} = 0$ for different
 masses.  
The solid lines describe
the subspace of separatrices originating at one of the minima while the dashed 
lines describe the subspace of separatrices ending up at another minimum.  }
\label{trspace}
\end{figure}

 Generally, two 1--dimensional lines embedded in the 2--dimensional space
 (the space of all trajectories with Im ${\cal W} \ =\ 0$) can cross at a finite
 number of points. What is this number and whether the lines cross at all
 depends on dynamics. We have found the presence of two intersections at
 $m < m_*$, of one intersection at $m = m_*$ and the absence of intersections
 at $m > m_*$. This picture illustrated schematically in Fig. \ref{trspace}
 looks quite natural.
\footnote{It is instructive to compare the situation with that in the model
analyzed recently in Ref.\cite{degen} where a family of degenerate BPS wall
solutions was observed. The model  involved two chiral superfields with
 the superpotential 
$$W\  =\ \frac {m^2}\lambda \Phi\ -\ \frac \lambda 3 \Phi^3\ -\
 \alpha \Phi X^2$$
BPS system involves here 4 singular points (4 vacua of the theory): two 
saddles, a stable and an unstable focuses. The BPS walls run from the instable
focus to the stable one. Again, the families of the relevant separatrices
are 1--dimensional. As in our case, the family of {\it all} trajectories is
3--dimensional. Here, however, the wall solutions lie in the one--dimensional 
subspace of the
{\it real}
trajectories satisfying  the two conditions ${\rm Im}\ \Phi = 0$ and
 ${\rm Im}\
X = 0$ (all the vacua are real here). Hence, the space of the BPS wall 
solutions is $ 1 + 1 - (3-2) \ =$  1--dimensional.}

\section{Wallsome Sphalerons.}
\begin{figure}
\grpicture{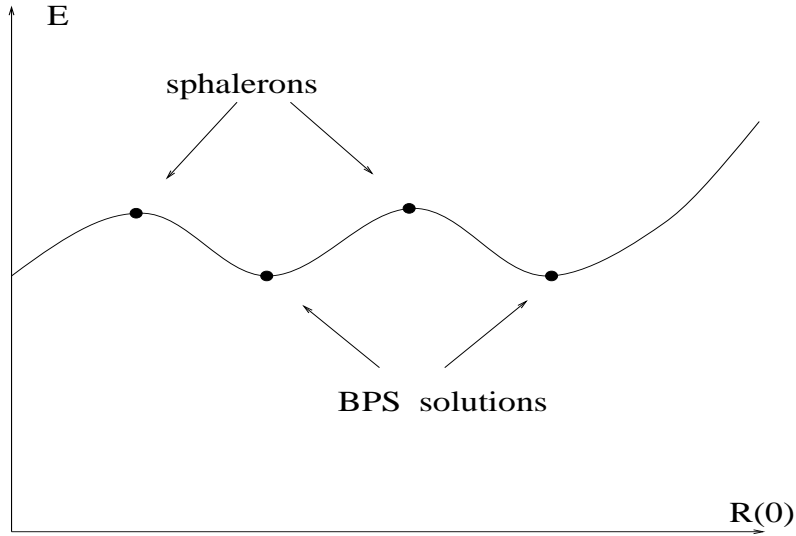}
\caption{Stationary points of the energy functional, $m < m_*$.}
\label{plER0}
\end{figure}

\setcounter{equation}0
The BPS solutions realize the absolute minimum of the energy functional
for all wall configurations. The other configurations have larger energy.
In the region $m < m_*$, we have found two different BPS solutions which
 should be separated by an energy barrier. Besides two nontrivial complex 
wall
 solutions, there is always a configuration describing two real walls
separated at infinite distance. The energy of such a configuration is twice
the energy of the real wall which coincides in our [$SU(2)$] case with the
 energy of the
 complex wall. There should be also an energy barrier separating this
 configuration from the lower BPS solution. This situation is illustrated
in Fig. \ref{plER0} where the energy functional minimized over all symmetric
wall configurations with a fixed value of $R(0)$ (the value of $|\phi|$ in
the middle) is drawn schematically as a function of $R(0)$. We see that
besides the minima (the BPS solutions), this plot has local maxima. In fact,
they are the saddle points of the energy functional with only one unstable
 mode corresponding to changing $R(0)$. 
Such saddle point configurations (the sphalerons) present non-trivial
 solutions
of the equations of motion. They are not BPS solutions, of course.

Thus, to study sphalerons, we should solve the full-scale second order
 equations of motion for the effective lagrangian with the potential
(\ref{potTVY})
\footnote{As usual, the sign of the potential should be reversed: we are
 looking for the minimum of the energy functional, 
not the action functional.}
and the standard kinetic term. Introducing again the polar variables
$\chi \ = \ \rho e^{i\alpha},\ \phi \ =\ Re^{i\beta}$, the equations can be
written in the form
  \beq
 \label{eqmot}
R'' - R \beta'^2 \ =\ 8R^3 [L(L + 3/2) + \beta_+^2] +
\frac{16R^5}{3\rho^2} - 4mR^2 \cos (\beta_-) \nonumber \\
R\beta'' + 2R'\beta'\  = \ 12R^3\beta_+ + 4mR^2 \sin(\beta_-) \nonumber \\
\rho'' - \rho \alpha'^2 \ =\ \frac{8R^4}\rho L + m^2\rho - 
\frac{16R^6}{9\rho^3}  \nonumber \\
\rho \alpha'' + 2\rho' \alpha' \ =\ \frac{8R^4}{\rho} \beta_+ -
\frac{8mR^3}{3\rho} \sin (\beta_-)
  \eeq
where $L = \ln(R^3 \rho^2), \ \beta_\pm = 3\beta \pm 2\alpha$. The system
(\ref{eqmot}) involves one integral of motion
   \beq
 \label{cons}
T - U \ =\ R'^2 + R^2 \beta'^2 + \rho'^2 + \rho^2 \alpha'^2 - \nonumber \\
\left[ 4R^4 (L^2 + \beta_+^2)
+ m^2\rho^2 + \frac {16R^6}{9\rho^2} - \frac{8mR^3}3 \cos (\beta_-)
\right]  
\ =\ {\rm const}
  \eeq
Solving the system (\ref{eqmot}) is a more complicated technical problem than
that for the BPS system (\ref{4sys}). The Cauchy problem involves here
8 initial conditions. As earlier, we can restrict our search with the
symmetric configurations [see Eq.(\ref{sym})]. Then 4 initial conditions
are fixed:
$$ R'(0) = 0,\ \rho'(0) = 0,\ \alpha(0) = \pi/4, \ \beta(0) = -\pi/6$$
Four other initial values satisfy the relation (\ref{cons}) with {\it const}
 $= 0$. Thus, we are left with 3 free
parameters, say, $\rho(0),\ R(0)$, and $\beta'(0)$, which should be fitted
so that the solution approach the complex vacuum in Eq.(\ref{minn0}) at
$z \to \infty$. Technically, we used the mismatch parameter in
 Eq.(\ref{mism})
as a criterium, the initial data were fitted to minimize it.

\begin{figure}
\grpicture{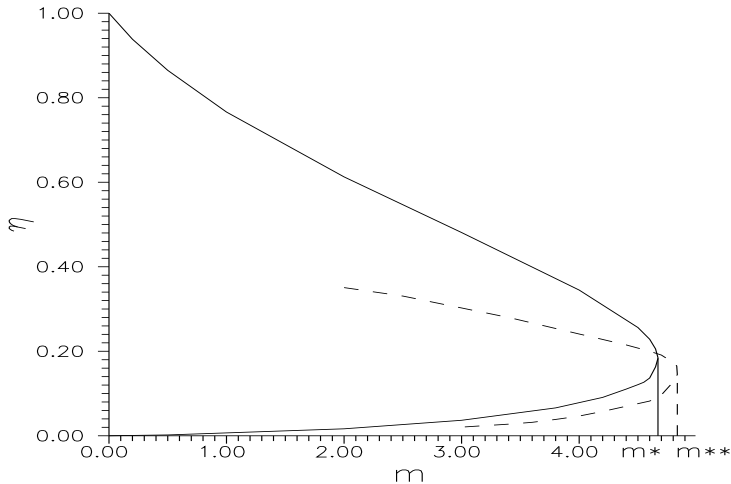}
\caption{The solutions of equations of motion at different masses. Solid lines
correspond to the BPS solutions and dashed lines --- to the sphalerons or 
non-BPS local minima.}
\label{vses}
\end{figure}

The results are presented in Fig. \ref{vses}. We obtain first of all our old
BPS solutions (solid lines in Fig.\ref{vses}). We find also two new 
 solution branches drawn with the dashed lines in Fig. \ref{vses}. We see
 that,
similarly to the BPS branches, two new dashed branches fuse together at some
$m = m_{**} \approx 4.83$.
\footnote{To be on the safe side, we would claim $m_{**} = 4.835 \pm .020$. 
The accuracy here is much lower than that for $m_*$ precisely because looking
for the minimum of $\Delta$ in the space of 3 parameters is much more 
complicated numerical problem than fitting just 1 parameter which we did
 while searching for BPS solutions. That is also the reason why the dashed
 branches are not drawn in the region of small masses. Our accuracy is not
sufficient there. }
 No solution for the system (\ref{eqmot}) exists at
$m > m_{**}$. For large masses, the minimal mismatch parameter $\Delta$ 
defined in
Eq.(\ref{mism}) is not zero, and the larger is $m - m_{**}$, the larger 
$\Delta$ is. Numerically, we find in this case
 \beq
 \label{Deleqmot}
\Delta_{\rm eq. mot.}(m) \ =\ 0.052 (m-m_{**})^{0.82}
  \eeq
in the full analogy with Eq.(\ref{DelBPS}).

\begin{figure}
\grpicture{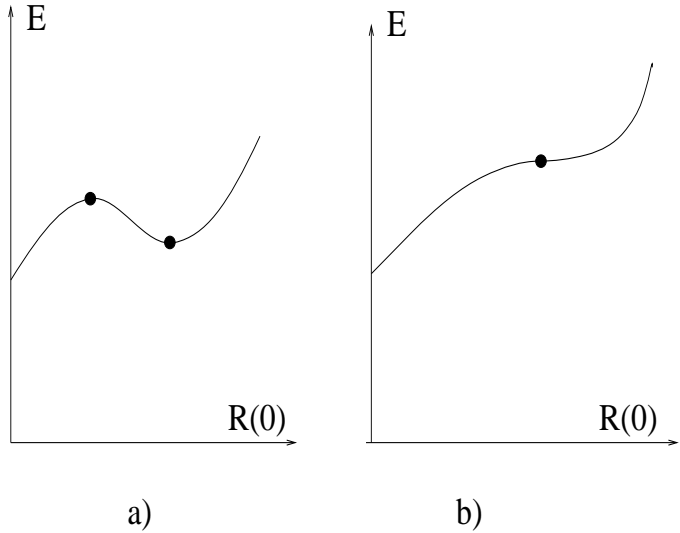}
\caption{Stationary points of the energy functional:
 {\it a)} $m_* < m < m_{**}$, {\it b)} $ m = m_{**}$.}
\label{plER0h}
\end{figure}

The physical interpretation of these new solutions is rather transparent. For
$m < m_*$, they present the sphalerons as shown in Fig. \ref{plER0}. At 
$m = m_*$,
two  BPS minima fuse together and the energy barrier separating them 
disappears. The upper sphaleron branch meets 
with the BPS branches at this point. When $m$ is slightly above $m_*$, the
former BPS minimum  is still a local minimum of the energy functional, but 
its
 energy is now slightly above the BPS bound (see Fig. \ref{plER0h}a). 
The corresponding solution is described by the analytic continuation of the 
upper sphaleron branch. The lower dashed branch in the region 
$m_* < m < m_{**}$ is still a sphaleron. At the second critical point
$m = m_{**}$, the picture is changed again (see Fig. \ref{plER0h}b). The local
maximum and the local minimum fuse together and 
the only one remaining stationary point does not correspond to an extremum of
 the energy 
functional anymore. At larger masses, no non-trivial stationary points are
left.

Like in the simpler BPS case, the presence of the discrete number of the
wall solutions for the system (\ref{eqmot}) depending on the value of the 
parameter $m$ can be naturally interpreted in the language of the qualitative
theory of differential equations. The phase space is now 8--dimensional and
the space of all {\it trajectories} is 7--dimensional. There is one integral
of motion (\ref{cons}) and our wall solutions belong to the 6--dimensional
subspace of the trajectories with $T - U \ =\ 0$. Again, the minima 
(\ref{minn0})
with zero derivatives $\dot\phi \ =\ \dot\chi\ =\ 0$ present the critical 
points of the equation system (\ref{eqmot}). The corresponding $8 \times 8$
Jacobi matrices involve 4 positive and 4 negative eigenvalues [the 
eigenvalues are actually the same as for the BPS system (\ref{4sys}), but 
each eigenvalue is now double degenerate]. Thereby, the family of 
separatrices which go down from one of the critical points is 
3--dimensional. The same is true for the family of separatrices which end 
up at the other minimum. Generally, two 3--dimensional manifolds embedded 
in the 6--dimensional space of all trajectories with $T - U\ =\ 0$ may 
cross at a finite number of points. We have seen that the crossing occurs at
4,2,1, or 0 points depending on the value of mass. This looks 
reasonable though in this case we are not clever enough to visualize it.

We hasten to comment that the height of the energy barrier as shown in Figs.
\ref{plER0}, \ref{plER0h}  is very much exaggerated. In fact, we could not
even accurately determine the difference of the energy of the lower sphaleron
 branch
and of all the dashed branches in the intermediate mass region
 $m_* < m < m_{**}$
from the BPS bound $\epsilon_{BPS}$. We can only say that the relative
 difference $\kappa = (\epsilon - \epsilon_{BPS})/\epsilon_{BPS}   $
is well below $10^{-4}$ in this case. The appearance
 out of the blue of such a small number which  is not
 present neither in supersymmetric QCD lagrangian
 nor in the effective TVY theory is rather surprising. Right now we do not see
a profound theoretical reason for this ``experimental fact'', but the numerical
evidence is unequivocal. The fact that the barrier is very low conforms well
with the fact that the second phase transition occurs pretty soon after the
first one: the values $m_* \approx 4.67$ and $m_{**} \approx 4.83$ are very
close.

\begin{figure}
\grpicture{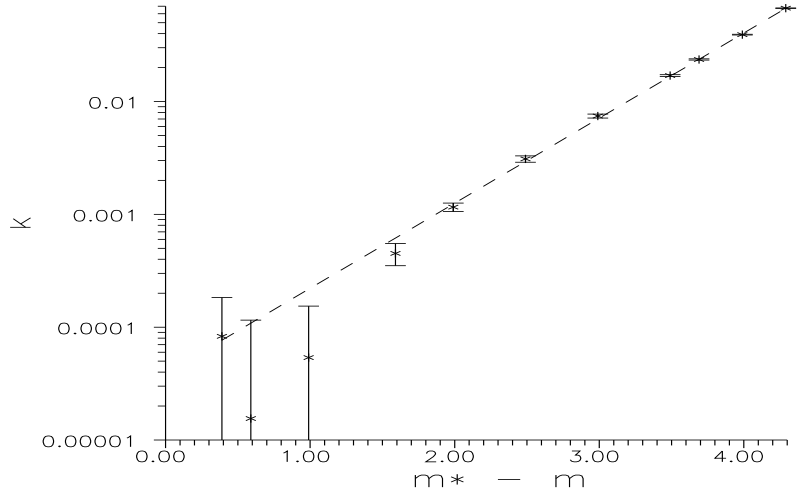}
\caption{Energy of the upper sphaleron as a function of mass.}
\label{Esph}
\end{figure}

The difference of the energy of the upper sphaleron from the BPS bound can be
numerically evaluated in a certain mass region. The results are plotted
in Fig. \ref{Esph}. Naturally, the height of the barrier goes to zero when
$m$ approaches $m_*$. But, as seen from  Fig. \ref{Esph}, even at small masses,
the barrier height is still rather small. In the range of $m_* - m$ where we
could calculate $\kappa$ reliably, it follows nicely the fit
$\kappa \approx \ 3.90 \cdot 10^{-5} \exp \{1.74 (m_* - m)\}$ (the dashed line
 in Fig. \ref{Esph}) 

\begin{figure}
\grpicture{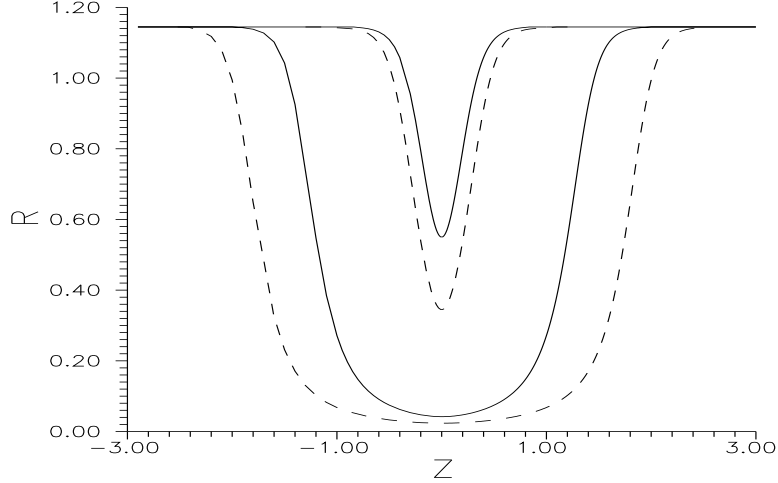}
\caption{Profiles $R(z)$ for BPS walls (solid lines) and wallsome 
sphalerons (dashed lines) for $m=3$.}
\label{prof}
\end{figure}

Finally, we present for illustration the profiles $R(z)$ for all four
branches at $m = 3$ (see Fig. \ref{prof}).

\section{Discussion.}
\setcounter{equation}0 

We have found that the properties of the system are
drastically changed at $m = 4.67059\ldots \Lambda$ and further at
$m \approx 4.83 \Lambda$. It makes sense to
express the result in invariant terms and to trade $\Lambda$ for an                    
invariant physical quantity such as the gluino condensate $\Sigma = |
<{\rm Tr}\ \lambda^2>| $ in a chirally asymmetric vacuum. From
Eqs.(\ref{minn0}), (\ref{norm}), we obtain 
   \beq
\Sigma \ =\ \frac{16\pi^2}{\sqrt{3}} m^{1/2} \Lambda^{5/2}
\label{SigLam}
  \eeq
 Thus, the phase transitions occur at
 \beq
 \label{mSig}
m_* \ \approx\  m_{**} \ \approx \ 0.8 \Sigma^{1/3}
 \eeq

Needless to say, they are not  phase transitions of a habitual
 thermodynamic variety. In particular, the vacuum energy is zero both
  below and above the phase transition point --- supersymmetry is never
  broken here. Hence $E_{vac} (m) \equiv 0$ is not singular at 
$m = m_*, m_{**}$.
  Some similarities may be observed with the 2--dimensional Sine--Gordon
  model where the number of the states in the spectrum depends on the 
  coupling constant $\beta$ so that the states appear or disappear at some
  critical values of $\beta$ \cite{SG}. May be a more close analogy can be
  drawn with the $N=2$ supersymmetric Yang--Mills theory. The spectrum of
that system depends on the Higgs expectation value 
$u\  = \ <{\rm Tr} \ \varphi^2>$.
A study of the exact solution of the model due to Seiberg and Witten 
\cite{SW} displays the existence of a ``marginal stability curve'' in
the complex $u$--plane \cite{stab}. When crossing this curve, the spectrum
pattern is qualitatively changed.

The particular numerical value (\ref{mSig}) was obtained by studying the
effective TVY lagrangian with the superpotential (\ref{WTVY}) and 
the standard kinetic term  $|\partial \phi|^2 + |\partial \chi|^2$ which
follows from the term $\frac 14 \int d^4\theta (\bar\Phi \Phi + \bar X X)$
in the superlagrangian.
However, in contrast to the superpotential, the form of the kinetic term in
 the effective lagrangian cannot be fixed from the symmetry considerations
alone. As as obvious generalization, one can write
  \beq
 \label{Lkin}
{\cal L}_{\rm kin} \ =\ \frac \alpha 4 \int d^4\theta \bar \Phi \Phi +
\frac \beta 4 \int d^4\theta \bar X X
  \eeq
Adding the term $\frac 12 \int {\cal W}(\Phi, X) d^2\theta \ +$ H.c. and 
excluding the
 auxiliary fields, the bosonic part of the lagrangian acquires the form
  \beq
 \label{Lab}
{\cal L} \ =\ \alpha |\partial \phi|^2 + \beta |\partial \chi|^2
- \frac 1 \alpha \left|\frac {\partial {\cal W}}{\partial \phi} \right|^2
 - \frac 1 \beta \left|\frac {\partial {\cal W}}{\partial \chi} \right|^2
  \eeq
The BPS equations are modified to 
   \beq
 \alpha \frac{ \partial \phi}{\partial z} \ =\ \pm \frac 
{\partial \bar {\cal W}} {\partial \bar \phi},
  \ \ \ \ \     \beta
\frac{ \partial \chi}{\partial z} \ =\ \pm \frac 
{\partial \bar {\cal W}} {\partial \bar \chi},
  \label{BPSab}
  \eeq
Suppose first that $\alpha = \beta$. Then the equations (\ref{BPSab}) can
be reduced in the standard form (\ref{BPS}) by rescaling the independent
variable $z  = \alpha z'$. The walls become thicker (or thinner) but, in all
other respects, the form and the properties of the solutions are the same.
If $\alpha \neq \beta$, the situation is not so trivial. The corresponding 
effective lagrangian seems to describe a different dynamics than that with 
$\alpha = \beta = 1$. Nevertheless, we will shortly see that the lagrangian
(\ref{Lab}) can still be reduced to the lagrangian with the standard kinetic
term by rescaling {\it both} the independent variable $z$ {\it and} the mass
parameter $m$.

Suppose, we have already set $\alpha = 1$ by rescaling z in a proper way. It is
not difficult to see that the BPS system (\ref{BPSab}) is reduced to the
standard system (\ref{BPS}) with the superpotential
  \beq
 \label{Wab}
{\cal W} \ =\ \frac 23 \Phi^3 \left[ \ln \frac{\Phi^3 X'^2}{\beta \Lambda^5} -
 1 \right] - \frac m{2\beta} X'^2
  \eeq
where $X' = \beta^{1/2} X$. The expression (\ref{Wab}) has the same functional
form as Eq.(\ref{WTVY}), only the scales for $\Lambda$ and $m$ have changed. 
Again, the BPS solutions disappear when $m > m_*(\beta) = 4.67059\ldots 
\beta^{6/5}$. In physical units, $m_*(\beta) \ \approx \ 
0.8 \beta \Sigma^{1/3}$. The second critical point scales in the same way.

Thus,  we cannot
claim that the phase transition in the theory of interest (\ref{LSQCD})
would occurs exactly at the  mass (\ref{mSig}). Our guess, however, is that
the true values of $m_*$ and $m_{**}$ in the supersymmetric QCD are
 close to the estimate (\ref{mSig}).
\footnote{Of course, the modification (\ref{Lkin}) is not the only possible 
one. One could invent more complicated kinetic terms involving higher 
(covariant)
derivatives and some fractional powers of $\Phi$ and $X$. We did not study 
these exotic modifications and cannot exclude a logical possibility that, 
after adding the terms with high derivatives, physics would dramatically change: e.g. the complex BPS walls would exist at any value of mass or would not 
exist at all. We do not believe, however, that this possibility is realized.}

To summarize, our main physical conclusion is the absence of complex domain
walls connecting different chirally asymmetric vacua in the supersymmetric
QCD at large enough masses. In particular, they are absent in the limit
$m \to \infty$, for the pure supersymmetric Yang--Mills theory.

In recent \cite{Witten}, the complex domain walls in the pure $N=1$ SYM theory 
were  discussed  in the context of D--brane dynamics. It was shown, in 
particular, that these walls, in spite of the fact that, on the fundamental
level, they are composed
of the fundamental gluon and gluino fields in the adjoint representation
of the gauge group, behave like  objects in the fundamental representation
in color. In particular, a QCD string originating at a heavy fundamental source
may end up at the domain wall (cf. the analogous situation in the massless
adjoint $QCD_2$ where fundamental colored sources are screened in spite of the
 fact that the lagrangian of the theory does not involve the fields which
are fundamental in color \cite{Kleb}). But that was in the {\it assumption} 
that such walls are there which, as we have seen, is not true.

The walls interpolating between a chirally asymmetric and the chirally
symmetric vacua are present at any value of mass and also in the limit
$m \to \infty$. For the time being, it is difficult for us to say something
about their color properties: their very existence has been detected only
in the framework of the effective TVY lagrangian involving only colorless
composite fields. 

Finally, in the light of our new results, let us discuss  the problem of 
torons. Torons are Euclidean gauge field configurations with fractional
topological charge (\ref{nu}), $\nu$ being the integer multiple of $1/N$, 
which
may in principle contribute in the Euclidean path integral in the theory
involving only adjoint color fields regularized in the infrared by placing it
 in a finite 4--dimensional toroidal box \cite{tor}.  Whether
such toron contributions are relevant or not in the supersymmetric gluodynamics
 is an old and controversial question (see e.g.
\cite{tordis}). It was discussed in details recently in \cite{my}, and we will 
not present here anew all the arguments {\it pro} and {\it contra}.

We would like only to reiterate here that the presence of torons would 
substantially modify the physical picture for the vacuum structure of the
supersymmetric gluodynamics (vis. $N+1$ degenerate vacua and $N$ different
real domain walls interpolating between them) advocated in \cite{my} and in
the present paper. In the first place, the presence of torons would modify
the gluing prescription for the effective potential (\ref{potVY}). The gluing 
would occur at ${\rm arg} (\phi^3) = \pi$ rather than at 
 ${\rm arg} (\phi^3) = \pi/2$ and  ${\rm arg} (\phi^3) = 3\pi/2$ as is the case
when only the instanton--like gauge field configurations with integer 
topological charge are taken into account. Then the potential (\ref{potVY}) 
would have only one non--trivial minimum at  ${\rm arg} (\phi^3) = 0$, not two
such minima as before. The theory would involve only two vacua, the chirally
asymmetric and the chirally symmetric one, and  only one interpolating 
domain wall. 

We have seen, however, that, at least if the supersymmetric gluodynamics is 
understood as the supersymmetric $QCD$ in the limit $m \to \infty$, this
{\it is} not true. When the matter chiral multiplets  are added, torons just
do not exist however large the mass is (the Dirac equation for fundamental
 color
fields does not make sense on a toron background), and the picture with 3
[$N+1$ for the $SU(N)$ case] vacuum states {\it is} correct.

A nontrivial feature of the theory under consideration is that the {\it real}
domain walls do not decouple in the limit $m \to \infty$. That is in contrast
to the simple two--dimensional model discussed in Ref.\cite{my} where the
energy of a 
two--dimensional {\it analogue}  of the 4--dimensional supersymmetric domain
walls becomes infinite in the limit $m \to \infty$ so that, in this
limit, the vacua decouple completely from each other, and the physical
picture corresponds not to the spontaneous breaking of a discrete symmetry,
 but to the appearance of a new {\it superselection rule} in the theory
which just corresponds to the {\it analogue} of toron contributions in the
path integral being switched on. 

Note further that, for the supersymmetric $QCD$ with large mass, the problem
of how the different branches of logarithm in the potential are glued together
is completely irrelevant as far as the wall solutions are concerned. For the
real walls, we have always ${\rm arg} (\phi^3 \chi^2) = 0$ irrespectively
of whether ${\rm arg} (\phi^3) = 0$ (for the wall connecting
the vacuum with the positive value of $<{\rm Tr}\  \lambda^2>$ with the
 chirally
symmetric one) or  ${\rm arg} (\phi^3) = \pi$ ( for the wall connecting the 
vacua
with $<{\rm Tr}\ \lambda^2>\  < 0$ and $<{\rm Tr}\ \lambda^2>\  = 0$). The phase 
of the 
heavy field $\chi$ always compensates the phase of $\phi$. In a sense, heavy
fields do not decouple {\it completely} in this case.
 
The question of whether it still makes sense to formulate the pure 
supersymmetric gluodynamics {\it not} as the large mass limit of 
supersymmetric $QCD$ but in some other way which takes into consideration
the toron contributions so that the theory would involve only {\it one}
domain wall connecting {\it two} different vacua is still open. To clarify
it, one should first try to understand better the dynamics of the chirally
symmetric phase.

In fact, the existence of the chirally symmetric vacuum is probably the
most interesting and unexpected result of the effective lagrangian approach.
What is especially surprising is that it has the same (zero) energy as the
chirally asymmetric vacua. Such degeneracy is not required by any symmetry
considerations. Note that the degeneracy is there only for the theory defined
at infinite volume. At small volumes when the effective gauge coupling is
small, the theory can be analyzed directly in terms of the fundamental gluon
and gluino fields, there are only N vacuum states with nonzero value of the
 condensate, and no trace of the chirally symmetric phase is seen 
\cite{Witind}. By analyticity, that should be true also for large (but finite)
volumes. Indeed, playing around with the finite volume formulas of
 Ref.\cite{Kovner} displays that the degeneracy is lifted also for large 
volumes, the energy gap going to zero exponentially fast in the limit
 $L \to \infty$. 

There is a striking analogy with a two--dimensional model [$QCD_2$ with the
fermions in the adjoint color representation for higher unitary groups
 $SU(N \geq 3]$)
 studied in Ref.\cite{QCD2}. In this model,
the bosonized lagrangian displays the existence of several vacuum states 
 separated presumably by the domain walls (the latter are just solitons in two
dimensions). Like in the supersymmetric $QCD_4$, different vacua are 
distinguished by the different values of the fermion condensate
$<{\rm Tr}\ \{\lambda_R \lambda_L\}>$. For higher even $N$, the existence of 
several vacua does not follow from symmetry considerations (for odd
$N$, it {\it does}). That was formulated as a paradox in Ref.\cite{QCD2}
but, bearing in mind fresh insights from supersymmetric $QCD_4$, it looks
now less surprising.

\vspace{.5cm}

{\bf Acknowledgments}: \hspace{0.2cm} 
We are indebted to M. Shifman for useful comments and 
to N. Koniukhova for the interest and the  advices 
concerning the numerical aspect of the problem.
 A.S. acknowledges warm hospitality extended to him at Saclay where this work
was finished.
This work was supported in part  by the RFBR--INTAS grants 93--0283, 94--2851,
and 95--0681, by the RFFI grants 96--02--17230,  97--02--17491, and 
97--02--16131, by the RFBR--DRF grant 96--02--00088,
by the U.S. Civilian Research and Development Foundation under award 
\# RP2--132, and by the Schweizerishcher National 
Fonds grant \# 7SUPJ048716.

\end{document}